\documentclass[aps,prl,twocolumn]{revtex4}
\usepackage{graphicx}
\usepackage{bm}
\begin{document}
\title{Two-dimensional elastic turbulence}
\author{S. Berti$^1$, A. Bistagnino$^2$, G. Boffetta$^{2,4}$, 
A. Celani$^3$ and S. Musacchio$^2$}
\affiliation{
$^1$ Department of Mathematics and Statistics, University of Helsinki,
P.O. Box 68, FIN-00014 Helsinki, Finland \\
$^2$ Dipartimento di Fisica Generale and INFN, 
Universit\`a degli Studi di Torino, 
Via Pietro Giuria 1, 10125, Torino, Italy \\
$^3$ CNRS URA 2171, Institut Pasteur, 75724 Paris Cedex 15, France \\
$^4$ CNLS, LANL, Los Alamos, NM 87545, USA }
\date{\today}
\begin{abstract}
We investigate the effect of polymer additives on a two-dimensional
Kolmogorov flow at very low Reynolds numbers by direct numerical 
simulations of the Oldroyd-B viscoelastic model. We find that
above the elastic instability
threshold the flow develops the elastic turbulence regime
recently observed in experiments.
We observe that both the turbulent drag and the Lyapunov exponent increase 
with Weissenberg, indicating the presence of a disordered, turbulent-like
mixing flow. The energy spectrum develops a power-law scaling range with
an exponent close to the experimental and theoretical expectations.
\end{abstract}
\maketitle

One of the most remarkable effects of highly viscous polymer solutions 
which has been recently observed in experiments is the development of
an ``elastic turbulence'' regime in the limit of 
strong elasticity \cite{GS00}. The flow of polymer solution
in this regime displays irregularities typical of turbulent
flows (broad range of active scales and growth of flow
resistance) even at low velocity and high viscosity, i.e. in the
limit of vanishing Reynolds number. As a consequence of 
turbulent motion at small scales, elastic turbulence has been
proposed as an efficient technique for mixing in very low Reynolds
flows, such as in microchannel flows \cite{GS01,BSBGS04,AVG05}.
Despite its great technological interest, elastic turbulence
is still only partially understood from a theoretical point of view.
Recent theoretical predictions are based on simplified versions of 
viscoelastic models and on the analogy with MHD equations \cite{BFL01,FL03}.

In this letter we investigate the phenomenology of elastic turbulence
in direct numerical simulation of polymer solutions in two dimensions.
Our main objective is to show that usual viscoelastic models,
developed for studying high Reynolds turbulent flows, are able to 
capture, in the limit of vanishing Reynolds numbers, the main 
phenomenology of elastic turbulence, i.e. irregular temporal behavior 
and spatially disordered flow.
Despite the important geometrical differences,
our numerical results are in remarkable agreement with experimental
observations of elastic turbulence: this suggests the possibility
to understand elastic turbulence on the basis of known viscoelastic
models.

To describe the dynamics of a dilute polymer solutions we
adopt the well known linear Oldroyd-B model \cite{BCAH87}
\begin{equation}
\partial_t {\bm u} + ({\bm u}\cdot{\bm \nabla}) {\bm u}
=-{\bm \nabla p} + \nu {\Delta} {\bm u} + \frac{2 \eta\,\nu}{\tau}
{\bm \nabla}\cdot{\bm \sigma} + {\bm f}
\label{eq:1}
\end{equation}
\begin{equation}
\partial_t {\bm \sigma} + ({\bm u}\cdot{\bm \nabla}) {\bm \sigma}
= ({\bm \nabla \bm u})^T \cdot {\bm \sigma} + {\bm \sigma} \cdot
({\bm \nabla \bm u})
-2\frac{({\bm \sigma}-{\bm 1})}{\tau}
\label{eq:2}
\end{equation}
where ${\bm u}$ is the incompressible velocity field
and the symmetric positive definite matrix  ${\bm \sigma}$ represents
the normalized conformation tensor of polymer molecules and ${\bm 1}$ 
is the unit tensor.
The solvent viscosity is denoted by $\nu$ and
$\eta$ is the zero-shear contribution of polymers to the total solution
viscosity $\nu_{t}=\nu(1+\eta)$ and is proportional to 
the polymer concentration.
In absence of flow, ${\bm u}=0$, polymers relax to the equilibrium
configuration and ${\bm \sigma}={\bm 1}$. The trace
$\textrm{tr}{\bm \sigma}$ is therefore a measure of polymer elongation.

The simplest geometrical setup that will prove useful to study the
elastic turbulence regime for viscoelastic flows
is the periodic Kolmogorov flow in two dimensions \cite{AM60}. 
With the forcing ${\bm f}=(F\cos(y/L),0)$, the system of 
equations (\ref{eq:1}-\ref{eq:2}) has a laminar Kolmogorov fixed 
point given by
\begin{eqnarray}
& & {\bm u}=(U_0\cos(y/L),0) \nonumber \\
& &  \label{eq:3} \\
& & {\bm \sigma}=
\left(
\begin{array}{cc}
1+{\tau}^2\, {U_0^2 \over 2 L^2}\sin^2{(y/L)} & 
-{\tau} {U_0 \over 2 L} \sin{(y/L)} \\
-{\tau} {U_0 \over 2 L} \sin{(y/L)}  & 1
\end{array}
\right) \nonumber
\end{eqnarray}
with $F=[\nu U_0 (1+\eta)]/L^2$ \cite{BCMPV05}. 
The laminar flow fixes a characteristic
scale $L$, velocity $U_0$ and time $T=L/U_0$. In terms of these variables,
we define the Reynolds number as $Re={U_0L \over \nu_t}$ and the
Weissenberg number as $Wi={\tau U_0 \over L}$. The ratio of these numbers
defines the elasticity of the flow $El=Wi/Re$.

It is well known that the Kolmogorov flow displays instability with
respect to large-scale perturbations, i.e. with wavelength much
larger than $L$. In the Newtonian case, the instability arises
at $Re_{c}=\sqrt{2}$ \cite{MS61}. 
At small Reynolds numbers, 
the presence of polymers can change the stability diagram of 
laminar flows \cite{Larson92,GS96} or induce elastic instabilities 
which are not present in Newtonian fluids \cite{LSM90,Shaqfeh96,BCMPV05}.
The Kolmogorov flow is no exception, and recent
analytical and numerical investigations have found the complete instability
diagram in the $Re$-$Wi$ plane \cite{BCMPV05}. 
For the purpose of the present work, we just have to recall 
that linear stability 
analysis shows that for sufficient large values of elasticity,
the Kolmogorov flow displays purely elastic instabilities, even at
vanishing Reynolds number (see Fig.~1 of \cite{BCMPV05}).
We remark that the fact that the original flow has rectilinear
streamlines does not exclude the onset of the elastic instability.
Above 
the elastic instability the flow can develop a disordered secondary flow 
which persists in the limit of vanishing Reynolds number and eventually 
leads to the elastic turbulence regime \cite{GS04}.

The equations of motion (\ref{eq:1},\ref{eq:2}) are integrated by 
means of a pseudo-spectral method implemented on a two-dimensional grid 
of size $L_0=2 \pi$ with periodic boundary conditions at resolution $512^2$.
Nmerical integrations of viscoelastic 
models are limited by Hadamard instabilities associated with the loss 
of positiveness of the conformation tensor \cite{SB95}. 
These instabilities are particularly important at high elasticity 
and limit the possibility to investigate the elastic turbulent regime 
by direct implementation of equations (\ref{eq:1},\ref{eq:2}). 
To overcome this problem, we have implemented an algorithm based on a 
Cholesky decomposition of the conformation matrix that ensures 
symmetry and positive definiteness \cite{VC03} and allows to reach
high elasticities.

\begin{figure}[hbt]
\includegraphics[draft=false,scale=0.7]{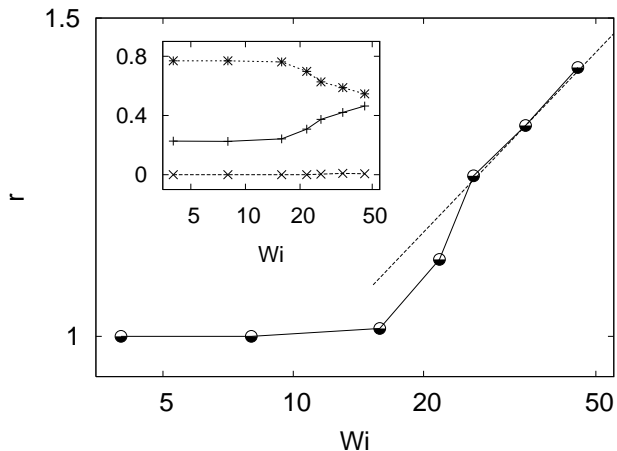}
\caption{
Mean power injection normalized with its laminar value
$r=P/P_{lam}$ as a function of  $Wi=\tau U/L$  for a set
of simulations with $\nu=0.769$, $\eta=0.3$, $L=1/4$ and $\tau=4$.
The elasticity is $El=64$ and the maximum Reynolds number
$Re=Wi/El$ is $Re = 0.7$. The line represents the power-law
behavior $r \sim Wi^{0.25}$.
Inset: amplitude of the Reynolds stress $\Pi_{r}$ ($\times$),
polymer stress $\Pi_{p}$ ($+$) and viscous stress $\Pi_{\nu}$ ($*$)
nondimensionalized by the total stress amplitude $FL$.
}
\label{fig1}
\end{figure}

One of the main features of a turbulent-like regime 
is the growth of the flow resistance to external forcing. 
This can be quantified as the power needed to maintain a given
mean velocity in the turbulent flow. 
The power injection in (\ref{eq:1}) is 
$P=\langle {\bf f} \cdot {\bf u} \rangle$
which, for the laminar flow (\ref{eq:3}) becomes
$P_{lam}=U_0^2 \nu(1+\eta)/(2 L^2)$. 
A remarkable feature of the Kolmogorov flow is that even in the
turbulent regime the mean velocity and conformation tensor are
accurately described by sinusoidal profiles \cite{BCM05}:
$\langle u_x \rangle=U \cos(y/L)$,
$\langle \sigma_{xy} \rangle= -\Sigma \sin(y/L)$
with different amplitudes with respect to the laminar fixed point.
Therefore the reduced average power injection for the turbulent
flow is simply
\begin{equation}
r={P \over P_{lam}} = {F L^2 \over \nu(1+\eta) U}
\label{eq:4}
\end{equation}
Figure \ref{fig1} shows the behavior of the power injection as
a function of the Weissenberg number $Wi=\tau U/L$. 
We see that at $Wi \simeq 15$ there is a transition from the laminar
regime to a turbulent-like regime in which $r>1$.
The growth
for the higher values of $Wi$ is compatible with a power law
scaling $r \simeq Wi^{0.25}$ which is qualitatively similar
with experimental observations \cite{BSS07} where an exponent 
$0.49$ is found. 
The different exponent observed here can be ascribed to the two-dimensional
nature of our flow or to its geometrical property (rectilinear 
streamlines and absence of material boundaries).

Because the Reynolds number in Fig.\ref{fig1} is always small,
and therefore the inertial term in (\ref{eq:1}) is negligible,
it is natural to ask where is the origin of turbulent fluctuations.
The momentum budget, in stationary conditions, reads
\begin{equation}
\partial_y \Pi_{r} = \partial_y (\Pi_{\nu} + \Pi_{p}) + f_x
\label{eq:5}
\end{equation}
where $\Pi_{r}=\langle u_x u_y \rangle$ is the usual Reynolds stress,
$\Pi_{\nu}=\nu \partial_y \langle u_x \rangle$ the viscous stress
and $\Pi_{p}=2 \nu \eta \tau^{-1} \langle \sigma_{xy} \rangle$ is the
stress induced by polymers.
The numerical observation that also the Reynolds stress is well
described by a monochromatic profile, 
$\langle u_x u_y \rangle = U_2 \sin(y/L)$, allows us to write the
momentum budget for the amplitudes as
$F L = U_2 + \nu U/L + (2 \nu \eta/\tau)\Sigma$. The
inset of Fig.\ref{fig1} shows the different contributions 
(normalized with the total stress) as a function of $Wi$. In the laminar
regime ($Wi \to 0$) $U_2=0$ and from (\ref{eq:3}) one has 
$\Pi_{p}/\Pi_{\nu}=\eta$. Above the transition to elastic 
turbulence, the polymer stress starts growing and reaches a value 
close to the viscous stress at the present maximum Weissenberg number.
We remark that we observe no indication of saturation
and therefore we may expect $\Pi_{p}$ to become the dominant term
at larger values of $Wi$.
The contribution of the Reynolds stress always remains smaller
than $10^{-2}$, confirming the irrelevance of inertial terms.
This is the hallmark of elastic turbulence where
elastic stress has the role played by the Reynolds
stress in usual turbulence.

\begin{figure}[thb]
\includegraphics[draft=false,scale=0.23]{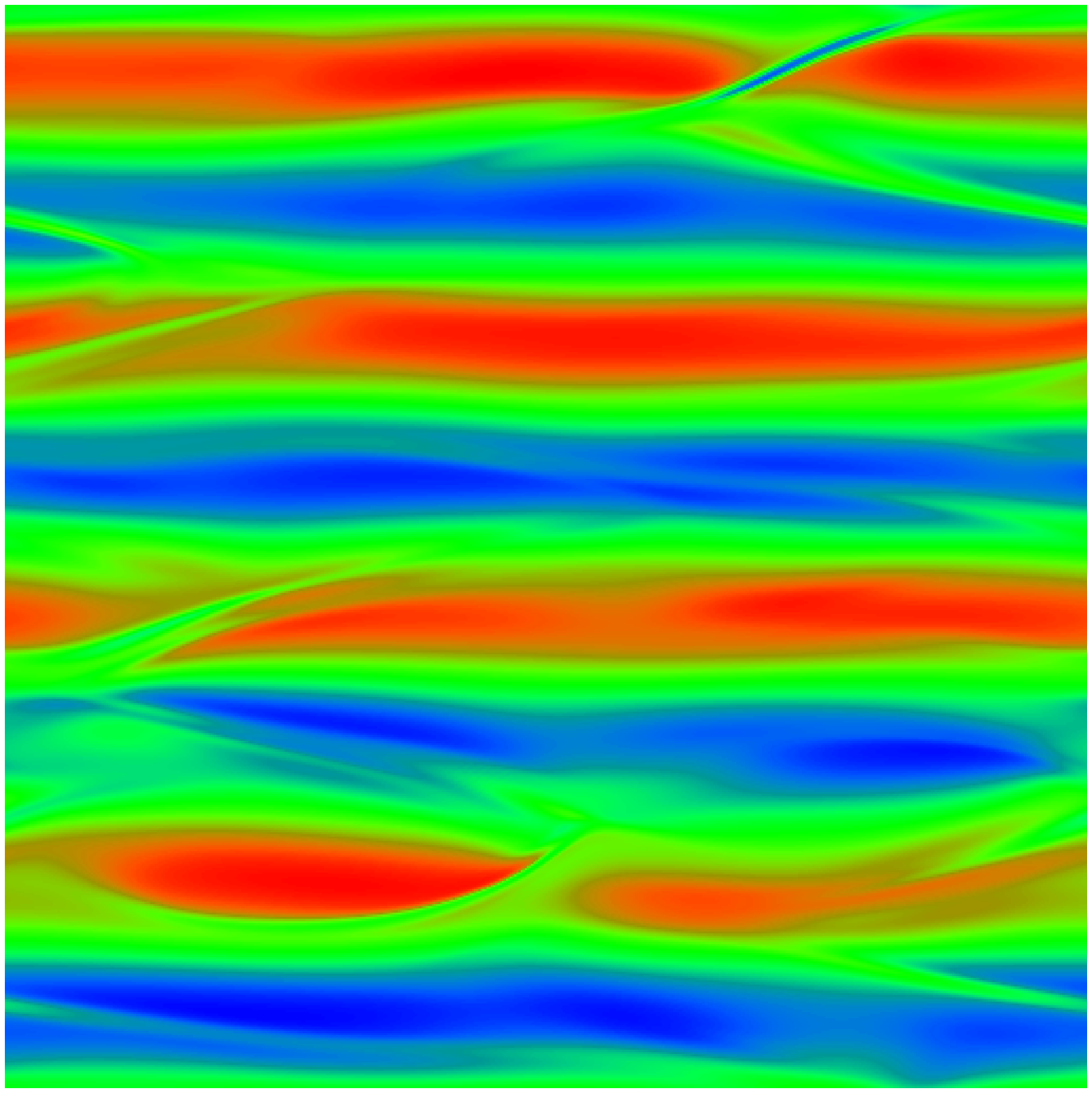}
\includegraphics[draft=false,scale=0.23]{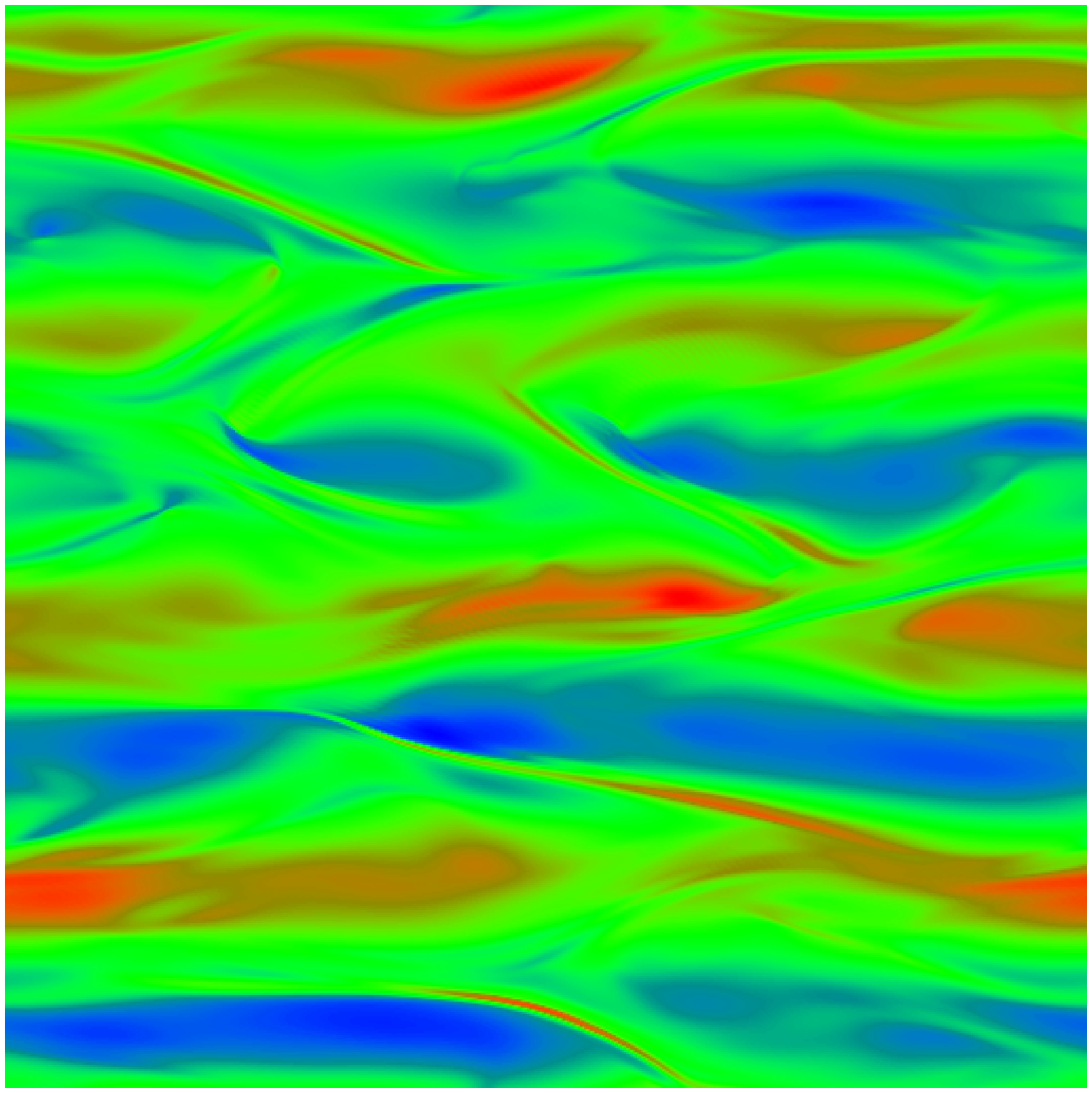}
\caption{(color online) Snapshot of vorticity field at 
$Wi=22$ (left) and $Wi=45$ (right). The flow is forced with
a Kolmogorov forcing $f_x=F cos(y/L)$ with $L=1/4$.}
\label{fig2}
\end{figure}

In order to get more insight in the elastic turbulence flow, 
in Fig.~\ref{fig2} we show two snapshots of the two-dimensional
vorticity field at two different $Wi$. The first snapshot is taken
at $Wi=22$, slightly above the elastic instability threshold.
The flow in this regime is still not turbulent 
and a secondary flow in the form of
thin filaments is clearly observable. These small scale filaments,
moving along the $x$ direction, are elastic waves, reminiscent
of the Alfven waves propagating in presence of a large scale 
magnetic field in plasma. 
Indeed, the possibility of observe elastic waves in polymer solution
was theoretically predicted within a simplified uniaxial elastic
model \cite{FL03} which has strong formal analogies with MHD equations,
but they were never observed before.

\begin{figure}[hbt]
\includegraphics[draft=false,scale=0.7]{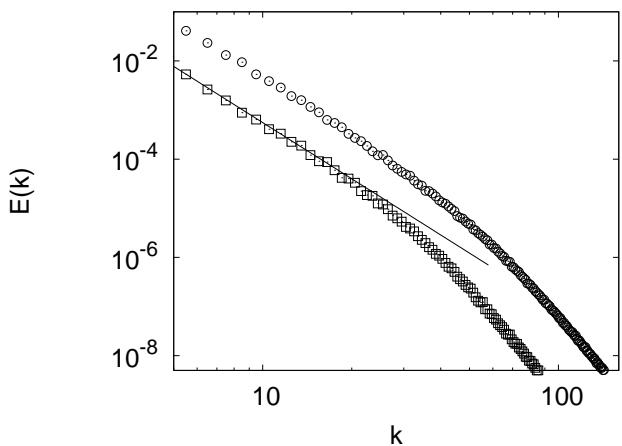}
\caption{Velocity fluctuation spectra at $Wi=26$ (squares)
and $Wi=45$ (circles). The line represents the
power-law behavior $k^{-3.8}$.}
\label{fig3}
\end{figure}

At higher values of elasticity the vorticity pattern becomes
progressively more irregular with chaotic motion of filaments. 
At $Wi=45$ we observe a highly irregular pattern
in which the underlying basic flow is 
hardly distinguishable.
This is the regime of elastic turbulence in which the flow
develops active modes at all the scales. Fig.~\ref{fig3}
shows the power spectrum of velocity fluctuations averaged 
over several configurations like the one shown in Fig.~\ref{fig2}. 
A power-law behavior $E(k) \simeq k^{-\alpha}$ is clearly observable 
with a spectral exponent $\alpha$ larger than $3$. 
Again, this is in quantitative agreement
with what observed in laboratory experiments \cite{GS00}
and with the theoretical predictions based on the uniaxial 
model \cite{FL03}.

One of the most promising applications of elastic turbulence is
efficient mixing at very low Reynolds number. This is an issue
of paramount importance in many industrial problems, namely in
microfluidic applications. Indeed, laboratory experiments 
in curvilinear channels have demonstrated that very viscous 
polymer solutions in the elastic turbulence regime are very 
efficient for small scale mixing \cite{GS01}. 
Mixing efficiency of polymer solutions has been 
studied in various setups, including microchannels \cite{BSBGS04}
and two-dimensional magnetically driven flows \cite{AVG05}.
Because in the elastic turbulent regime the flow is smooth
(i.e. the energy spectrum is steeper that $k^{-3}$)
a suitable characterization of mixing is given in terms 
of Lagrangian Lyapunov exponent $\lambda$ \cite{PV87}.
This is defined as the mean rate of separation of 
two infinitesimally close particles transported by the flow
and, in the present case, is related to the polymer stretching
rate \cite{BCM03}.

\begin{figure}[hbt]
\includegraphics[draft=false,scale=0.7]{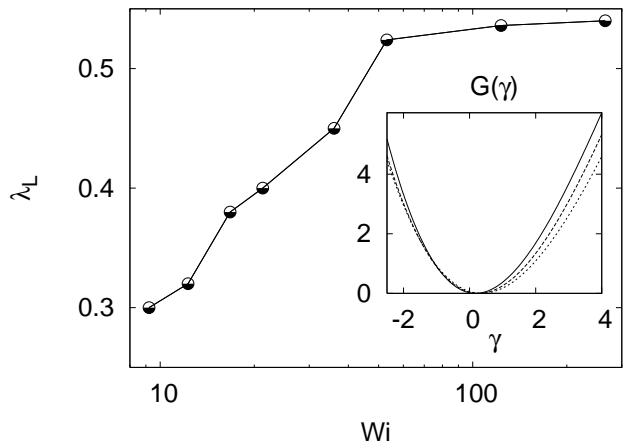}
\caption{Lagrangian Lyapunov exponent as a function of the Weissenberg 
number. Inset: Cramer function for three different Weissenberg numbers:
$Wi=12$ (solid line), $Wi=21$ (dashed line) and $Wi=53$ (dotted line).}
\label{fig4}
\end{figure}

Figure~\ref{fig4} shows the behavior of the Lyapunov exponent
as a function of $Wi$ at fixed $Re=1$. 
We observe that, above $Wi=10$, $\lambda$ grows  and saturates 
at values $\approx 2/\tau$ for $Wi$ larger than $60$.
We remark that this behavior is opposite to the one observed in 
the case of high Reynolds numbers
viscoelastic flows where the injection of
polymers reduces the degree of chaoticity by lowering $\lambda$ below 
$1/\tau$ \cite{BCM03}.

In the inset of Fig.~\ref{fig4} we plot the Cramer function
$G(\gamma)$ which is defined from the probability density functions
of finite-time Lyapunov exponents $P_t(\gamma)\sim \exp(-t G(\gamma))$
\cite{PV87}. As it is evident, increasing $Wi$ not only 
the degree of mean chaoticity increases, but also fluctuations becomes
larger, in particular the distribution of $\gamma$ becomes
asymmetric with a larger relative probability of positive fluctuations.
It is remarkable that the same qualitative behavior is 
observed in the case of high-Reynolds Newtonian turbulence, where
the distribution of Lyapunov fluctuations becomes more asymmetric 
with increasing $Re$ \cite{BBBCMT06}.
This suggests that in elastic turbulence elasticity (i.e. $Wi$) 
plays a similar role as non-linearity (i.e. $Re$) in ordinary hydrodynamic 
turbulence.

\begin{figure}[hbt]
\includegraphics[draft=false,scale=0.7]{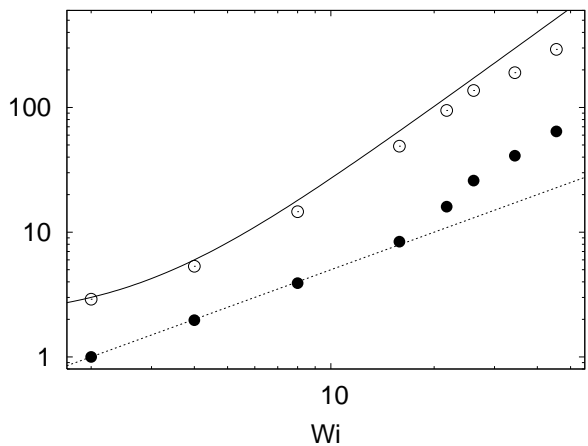}
\caption{
Average polymer elongation $\langle tr {\bm \sigma} \rangle$ (open
circles) and polymer stress amplitude $\Sigma$ (filled circles)
as a function of $Wi$. Lines represent the laminar behaviors
$2+Wi^2/4$ (continuous) and $Wi/2$ (dotted).}
\label{fig5}
\end{figure}

Finally, we have investigated the dependence of polymer statistics on 
the Weissenberg number. In Figure~\ref{fig5} we show the average 
squared polymer 
elongation $\langle \mathrm{tr} {\bm \sigma} \rangle$ integrated over 
the flow volume and the amplitude 
of cross stress $\Sigma$. At small $Wi$ these follow the laminar behavior,
i.e. $2+ Wi^2/4$ and  $Wi/2$, respectively. At the onset    
of elastic turbulence the cross polymer stress $\Sigma$
grows faster than linearly in 
$Wi$, as already shown in Figure~\ref{fig1}, and eventually appears to
approach a power-law behavior with a slope close to $1.5$.
On the contrary, the squared polymer elongation in elastic turbulence
grows more slowly than its laminar value. This is probably due
to the loss of coherence in stretching experienced in the turbulent flow.
At large $Wi$ an asymptotic behavior $ \sim Wi^{1.5}$ appears to set in
for $\mathrm{tr} {\bm \sigma}$ as well and the ratio 
$\langle tr {\bm \sigma} \rangle/\Sigma$ becomes constant.
 
Summarizing, we have shown that  elastic turbulence can be successfully 
reproduced numerically with the aid of a widely known viscoelastic model 
of polymer solutions (the Oldroyd-B model) and a simple geometrical setup
(the two-dimensional Kolmogorov flow). Most observed 
features have a strong qualitative 
resemblance with experimental results. Quantitative differences 
exist, however, and may be traced back to the 
two-dimensional or to the boundaryless
nature of our toy flow, or both. In this  context it would prove extremely 
useful to perform numerical simulations in more realistic geometries
and dimensionality to ascertain the origin of such differences. 

We thank V. Steinberg for valuable discussions. 
This work was carried out under the auspices of the National Nuclear
Security Administration of the U.S. Department of Energy at Los Alamos 
National Laboratory under Contract No. DE-AC52-06NA25396.
SB acknowledges the support of TEKES-2007 "Multimodel" project.



\begin{thebibliography}{99}


\bibitem{GS00}
A. Groisman and V. Steinberg, Nature {\bf 405}, 53 (2000). 

\bibitem{GS01}
A. Groisman and V. Steinberg, Nature {\bf 410}, 905 (2001). 

\bibitem{BSBGS04}
T. Burghelea, E. Segre, I. Bar-Joseph, A. Groisman and V. Steinberg, 
Phys. Rev. E {\bf 69}, 066305 (2004). 

\bibitem{AVG05}
P.E. Arratia, G.A. Voth and J.P. Gollub, 
Phys. Fluids {\bf 17} 053102 (2005). 

\bibitem{BFL01}
E. Balkovsky, A. Fouxon and V. Lebedev, 
Phys. Rev. E {\bf 64}, 056301 (2001). 

\bibitem{FL03}
A. Fouxon and V. Lebedev, 
Phys. Fluids {\bf 15}, 2060 (2003). 

\bibitem{BCAH87}
R.B. Bird, C.F. Curtiss, R.C. Armstrong, and O. Hassager, 
{\it Dynamics of polymeric fluids} Vol.2, Wiley, New York (1987). 

\bibitem{AM60}
V.I. Arnold and L. Meshalkin, Uspekhi Mat. Nauk {\bf 15}, 247 (1960).

\bibitem{BCMPV05}
G. Boffetta, A. Celani, A. Mazzino, A. Puliafito and M. Vergassola, 
J. Fluid Mech. {\bf 523}, 161 (2005). 

\bibitem{MS61}
L. Meshalkin and Y.G. Sinai J. Appl. Math. Mech. {\bf 25}, 1700 (1961). 

\bibitem{Larson92}
R.G. Larson, Rheol. Acta {\bf 31}, 213 (1992). 

\bibitem{GS96}
A. Groisman and V. Steinberg, Phys. Rev. Lett. {\bf 77}, 1480 (1996).

\bibitem{LSM90}
R.G. Larson, E.S.G. Shaqfeh and S.J. Muller, 
J. Fluid Mech. {\bf 218}, 573 (1990). 

\bibitem{Shaqfeh96}
E.S.G. Shaqfeh, Annu. Rev. Fluid Mech. {\bf 28}, 129 (1999). 

\bibitem{GS04}
A. Groisman and V. Steinberg, New J. Phys. {\bf 6}, 29 (2004). 

\bibitem{SB95}
R. Sureshkumar and A.N. Beris, 
J. Non-Newtonian Fluid Mech. {\bf 60}, 53 (1995). 

\bibitem{VC03}
T. Vaithianathan and L.R. Collins, J. Comp. Phys. {\bf 187}, 1 (2003). 

\bibitem{BCM05}
G. Boffetta, A. Celani and A. Mazzino, Phys. Rev. E {\bf 71}, 036307 (2005). 

\bibitem{BSS07}
T. Burghelea, E. Segre and V. Steinberg,
Phys. Fluids {\bf 19}, 053104 (2007).

\bibitem{PV87}
G. Paladin and A. Vulpiani, Phys. Rep. {\bf 156}, 147 (1987). 

\bibitem{BCM03}
G. Boffetta, A. Celani and S. Musacchio, Phys. Rev. Lett. 
{\bf 91}, 034501 (2003).

\bibitem{BBBCMT06}
J. Bec, L. Biferale, G. Boffetta, M. Cencini, S. Musacchio and F. Toschi,
Phys. Fluids {\bf 18}, 091702 (2006).

\end{thebibliography}
\end{document}